\documentclass[twocolumn,10pt,a4paper,sort&compress,aps,pra,showpacs,superscriptaddress]{revtex4-2}

\usepackage{hyperref}
\usepackage{graphicx}
\usepackage{xcolor}
\usepackage{dcolumn}
\usepackage{bm}
\usepackage{ulem}
\usepackage{color}
\usepackage{booktabs}
\usepackage{amsmath,amssymb,amsfonts}

\newcommand \bd{\color{black}}
\newcommand \revision{\color{black}}
\newcommand \cmedit{\color{black}}

\begin{document}

\preprint{APS/123-QED}

\title{The hydrodynamic Euler-elastica: shape transitions in the dynamical buckling of elastic filaments in Stokes flow}

\author{Cl\'{e}ment Moreau}
\email{clement.moreau@ls2n.fr}
\affiliation{Nantes Universit\'e, \'Ecole Centrale Nantes, CNRS, LS2N, UMR 6004, F-44000 Nantes, France}

\author{Laetitia Giraldi}
\email{laetitia.giraldi@inria.fr}
\affiliation{Universit\'{e} C\^{o}te d'Azur, Inria, CALISTO team, France}

\author{Hermes Bloomfield-Gad\^{e}lha}
\email{hermes.gadelha@bristol.ac.uk}
\affiliation{School of Engineering Mathematics and Technology and Bristol Robotics Laboratory, University of Bristol, UK}


\date{\today}

\begin{abstract}
The buckling of elastic filaments in viscous fluids, ubiquitous in biological systems like flagella, microtubules, and DNA, has long been described by the static Euler-elastica. 
Yet, when such filaments buckle dynamically, their shapes defy static predictions, exhibiting complex, unpredictable behaviours. 
Here, we use a coarse-grained numerical model to explore the long-timescale dynamics of filament buckling in Stokes flow, revealing three distinct morphological regimes, termed flip, loop, and knot.
The dominance of each regime is primarily governed by the dimensionless buckling number $\mathrm{Bu}$. Fourier analysis shows that these transitions between shape regimes arise from competition between the first three curvature modes, with high-order modes decay fitting an exponential law. 
In some parameter ranges, distinct shapes coexist for close initial conditions, indicating deterministic sensitivity to small perturbations.
These findings bridge static and dynamic buckling theories, with implications for biological propulsion and the design of microscale slender swimmers.
\end{abstract}

\maketitle

\section{\label{sec:intro}Introduction}

Flexible filaments are found everywhere in nature, particularly at the microscopic scale, where they are routinely subjected to large compressive loads or strong shear flows that trigger buckling instabilities \cite{tornberg2004simulating,liu2018morphological,becker2001instability,hall2019efficient}. Prominent examples include DNA strands \cite{fields2013euler,marko2012competition}, polymer chains \cite{deteresa1985model,golubovic2000flexible}, complex cytoskeletal microtubules \cite{schaap2006elastic} and actin filaments, as well as cilia and flagella \cite{vogel2012motor,gadelha2010nonlinear, cass2023reaction,ren2025swimming}. These latter structures play a central role in the locomotion of micro-organisms such as sperm cells \cite{gadelha2019flagellar,cass2023reaction}, diatom phytoplankton \cite{ardekani2017sedimentation} and bacteria propelling in fluids. Beyond biological systems, slender elastic structures undergoing buckling in viscosity-dominated environments are increasingly encountered in engineered contexts, including biomedical catheters \cite{hu2018steerable}, guidewires, and soft-robotic structures \cite{armanini2023soft,Yan2021MagnetoActive,Teo2025RoboSoft}, where controlled or uncontrolled buckling can critically affect performance and safety.

At steady-state, the shape of an elastic filament subjected to external forces is governed by the well-known Euler-elastica boundary value problem, which admits exact solutions in terms of elliptic functions \cite{truesdell1960rational,gadelha2018filament}. In this limit, contact forces balance exactly the imposed load, and the shape is defined by the torque balance \cite{fung2017classical,landau1986course,timoshenko2009theory,book:antman}. 
The Euler-elastica problem has been used as a basis to study numerous problems associated with the buckling of elastic filaments \cite{le2012buckling,blundell2009buckling}, with various modelling refinements allowing predictions of key behaviours in biological filaments. Notable examples include the incorporation of counterbend forces in filament bundles \cite{gadelha2018filament}, morphoelastic effects arising from growth or active internal stresses \cite{goldstein2006dynamic}, and the reaction--diffusion theory for motion patterns in active filaments \cite{cass2023reaction}, in which diffusion of an effective `bending' variable derived from the static Euler--elastica contribution can support self-organised pattern formation.

However, the Euler-elastica framework is limited to steady-state regimes. In contrast, the \textit{transient} dynamics of a filament buckling in a Stokes flow involve a continuously evolving balance between elastic forces, viscous drag, and time-dependent contact forces at the boundaries \cite{chopin2017dynamic,tornberg2004simulating, shelley2000stokesian}. The interplay between the decay of high-curvature modes, viscous resistance, and the forced convergence of endpoints leads to complex phenomena that often defy simple analytical descriptions.

The dynamical equations are typically formulated using Kirchhoff-Love theory \cite{fatehiboroujeni2021three, ling2018instability,de2017spontaneous,chelakkot2014flagellar,fily2020buckling} or Cosserat rod theory \cite{book:antman}, and hydrodynamic drag must be accounted for by using local or non-local approximations, such as resistive force theory \cite{gray1955propulsion} or variations of slender body theory \cite{Garg2022Cosserat,Maxian2023BendingFluctuations,Maxian2022Rotating}. The coupling between fluid-structure interaction and elasticity notoriously introduces computational stiffness in numerical resolution \cite{moreau2018asymptotic,fuchter2023three,tornberg2004simulating,shelley2000stokesian}. Buckling triggered by internal activity in flagella \cite{fily2020buckling,krishnamurthy2023emergent,cass2023reaction,oriola2017nonlinear} and unsteady shear flows \cite{bonacci2023dynamics,liu2018morphological} is known to yield chaotic behaviour. 
However, the long-timescale evolution of dynamic buckling, for set velocity boundary conditions, particularly the transition into post-buckling regimes, remains unexplored \cite{chopin2017dynamic,cammann2025form,du2019dynamics}.

While short-timescale dynamics have been investigated through the lens of unstable mode selection \cite{Wiggins1998a, chopin2017dynamic,lagrange2016wrinkling,gladden2005dynamic}, the long time-evolution and temporal behaviour where endpoints approach and eventually cross, beyond the static regimes \cite{gadelha2018filament} (Fig.~\ref{fig:elastica}), remains a significant gap in the literature. Most existing studies focus on the onset of instability and linear regimes or on static solutions. Moreover, to the best of our knowledge, none address the predictability of final shapes or the role of Fourier mode competition in determining the buckling instability outcome in passive elastohydrodynamic systems \cite{du2019dynamics}. 

Thus, inspired by the foundational Euler–elastica framework and its classical post-buckling solutions (Fig.~\ref{fig:elastica}) -- which describe the static evolution as endpoints approach and cross, producing the characteristic bent shapes and loop formation \cite{truesdell1960rational,matsutani2024euler} -- the present work aims to resolve the above gap in the literature by simulating the full dynamic buckling of a planar elastic filament embedded in a Stokesian fluid, driven at both ends at constant prescribed velocities towards each other (with hinged or clamped boundary conditions), from the onset of instability into the post-transient regime and through end-point crossing (Figs.~\ref{fig:snaps},~\ref{fig:clamp}). 
Unlike in the static elastica setting, the buckling here unfolds through transient elastohydrodynamic relaxation, so that the nonlinear development and final post-transient outcome depend on the elastohydrodynamic properties of the system and the actuation protocol, in particular the speed at which the endpoint is driven. 

A neighbouring line of work considers the flutter instability of filaments under follower forces (or follower-loads), in which a compressive load constrained to remain aligned with the local filament tangent can destabilise the straight state via a Hopf (flutter) bifurcation, leading to self-sustained oscillations and nonlinear post-bifurcation dynamics \cite{de2017spontaneous,clarke2024bifurcations,bayly2016steady}. This is connected with the present set-velocity actuation in that the end-contact force is an output of the model and can be decomposed into components along and normal to the local filament tangent; however, in our case the filament is driven by a kinematic constraint (the end-point position and speed are prescribed), whereas the follower-force model prescribes the load and allows the end point to evolve freely as the instability develops. While the overdamped Stokesian setting studied here does not support sustained oscillations, it complements the follower-force perspective by isolating how kinematic end actuation and viscous relaxation shape the nonlinear transient and post-transient evolution. Together, these viewpoints help clarify, in the spirit of the classical Euler–elastica framework, how tangentially directed compression coupled to geometry can promote self-organisation in viscous filament dynamics \cite{du2019dynamics}.

To this end, we build on and adapt the coarse-grained (CG) modelling framework introduced in \cite{moreau2018asymptotic} to investigate planar dynamic filament buckling in the Stokes regime, and report numerical observations that provide new insight into the viscous dynamical buckling of a filament. The CG method provides a robust and efficient reduced-order description while retaining the essential physics of elasticity and viscous dissipation \cite{Alouges2025Nlink,Mori2021Theoretical}. A key advantage is that it avoids solving for numerically stiff Lagrange multipliers (and associated boundary conditions) used to enforce filament inextensibility \cite{gadelha2010nonlinear,tornberg2004simulating}. The geometrically nonlinear elastohydrodynamic partial differential equations (PDEs) are thereby recast as a simpler system of ordinary differential equations (ODEs), and this approach has been generalised to a wide range of elastohydrodynamic problems \cite{fuchter2023three,walker2019filament,hall2019efficient,ntetsika2022numerical,cass2023reaction,cass2024predicting}, including experimental model construction on a robotic platform \cite{Teo2025RoboSoft}.

The paper is organised as follows. We first formulate the elastohydrodynamic model for a passive filament actuated at the distal end, considering different boundary conditions at the proximal end (from hinged to clamped), and introduce the nondimensionalisation. Using the coarse-grained (CG) framework \cite{moreau2018asymptotic}, we then compute the filament shape evolution together with the unknown contact forces at the endpoints. We benchmark the early evolution against static elastica solutions \cite{gadelha2018filament} by comparing shapes and horizontal forces, and identify clear departures between static and dynamic responses, highlighting the role of elasto-viscous buckling dynamics. We next show that the dynamics organise into distinct regimes controlled by a single nondimensional ‘buckling number’ $\mathrm{Bu}$, which captures the competition between filament elastohydrodynamic relaxation and endpoint motion timescales (set by the actuation speed). The main result is that the post-transient state exhibits three qualitatively different final shapes across ranges of $\mathrm{Bu}$: each dominates in a different regime, and multiple shapes can coexist near critical $\mathrm{Bu}$ values. In these coexistence regions, we observe bifurcations consistent with the onset of chaotic behaviour \cite{agrawal2022chaos,krishnamurthy2023emergent}. Finally, to connect these long-time outcomes to the onset of buckling, we analyse the filament dynamics by decomposing its curvature into Fourier modes, and model the decay of unstable modes with an exponential law.

\begin{figure}
    \centering
    \includegraphics[width=\columnwidth]{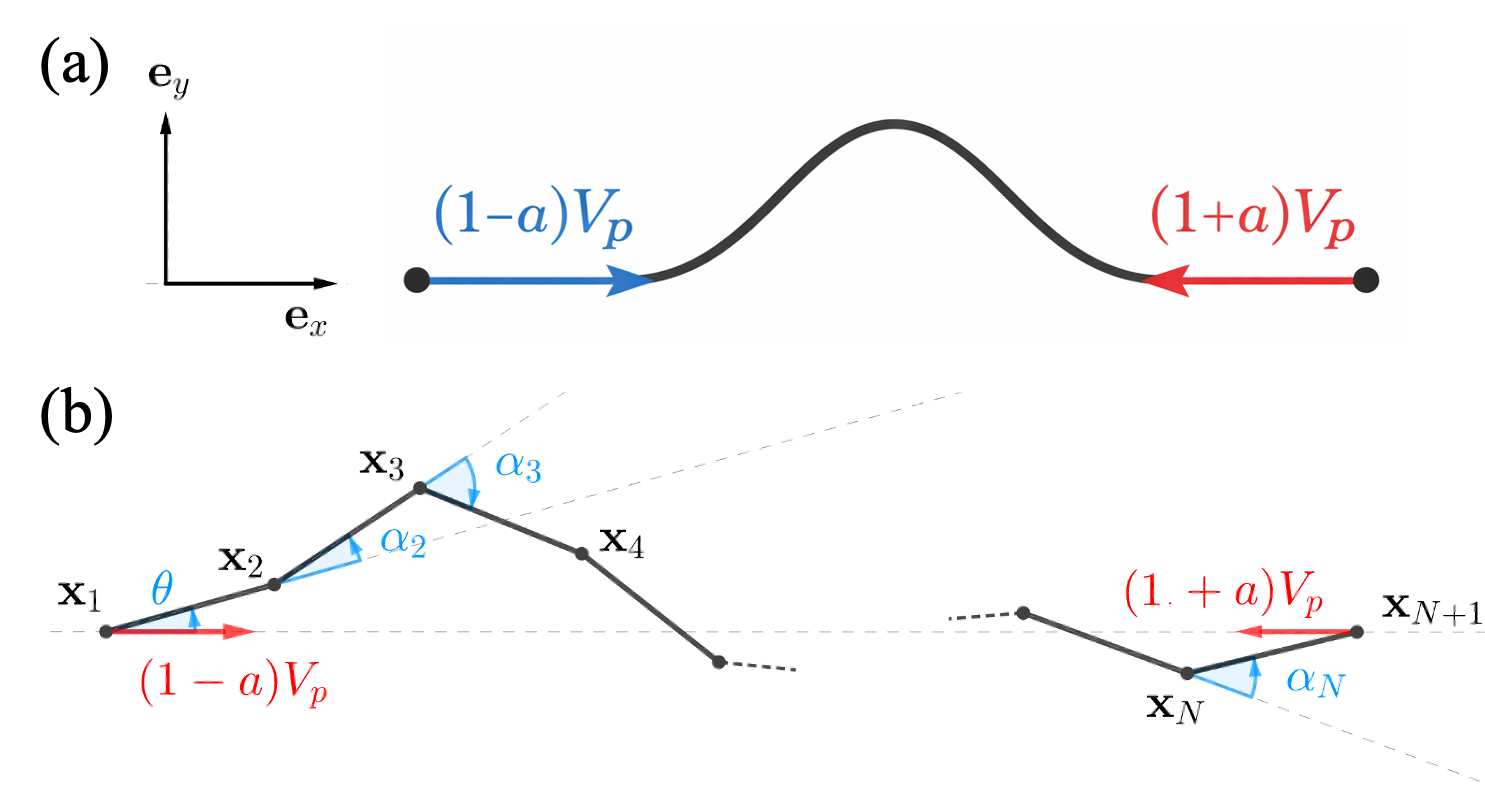}
    \caption{The hydrodynamic Euler-elastica set-up and coarse-grained representation of a filament driven at both ends. (a) Schematic of an inextensible planar filament in a Stokesian fluid, buckling under end-point actuation. The left and right end points are driven towards one another with prescribed velocities $(1-a)V_p\,\mathbf{e}_x$ and $-(1+a)V_p\,\mathbf{e}_x$, respectively, where $a$ partitions the imposed motion between the two ends. (b) Coarse-grained discretisation of the filament centreline into straight segments joining nodes $\mathbf{x}_i$, with local segment orientations $\alpha_i$ (blue) and end-point velocities shown in red. The filament shape evolves through the coupled dynamics of the node positions and segment angles \cite{moreau2018asymptotic}.}
    \label{fig:parameter}
\end{figure}

\section{The hydrodynamic Euler-elastica}
\label{sec:model}

We consider the motion of a flexible slender filament of length $L$ and bending stiffness $E_b$ in the reference plane $(\mathbf{e}_x,\mathbf{e}_y)$ (Fig. \ref{fig:parameter}).  
The filament is immersed in a Stokes flow (zero Reynolds number) and the hydrodynamic interaction is modelled using Resistive Force Theory \cite{gray1955propulsion}, approximating the hydrodynamic drag by a local linear anisotropic operator characterised by parameters $\eta_{\parallel}$ and $\eta_{\perp}$. Building on the coarse-grained formulation introduced in \cite{moreau2018asymptotic}, we approximate the filament with $N$ rod-like elements of length $\Delta s = L/N$. The filament configuration is thus parameterised by a vector of dimension $(N+2)$: the position of the proximal end $\mathbf{x}_1=(x_1,y_1)$, its orientation $\theta$ with respect to the reference axis $\mathbf{e}_x$, and the angles between consecutive elements, namely $\alpha_2 \dots \alpha_N$, as displayed on Fig. \ref{fig:parameter}.

Following \cite{moreau2018asymptotic}, the integration of the pointwise hydrodynamic force and torque density over the $i$-th segment yields the hydrodynamic force $\mathbf{f}^{\mathrm{h}}_i$ and moment $\mathbf{m}^{\mathrm{h}}_i$ exerted on each segment. 
To account for the filament's bending stiffness, we consider an internal elasticity torque $\mathbf{m}^{\mathbf{el}}_i$ at the junction between the $i$-th and $(i+1)$-th segments, and modelled as an Euler-Bernoulli rotational linkage where the torque is linearly related to the discrete curvature $\alpha_i$: 
\[ \mathbf{m}^{\mathbf{el}}_i= \frac{E_b}{\Delta s} \alpha_i \mathbf{e}_z. \]

At the proximal boundary, inspired by cantilevers and flagella attached to microswimmers' bodies, we may add a clamping term, modelled by a torque $\mathbf{m}_c$ applied at $\mathbf{x}_1$, proportional to the orientation $\theta$ of the first element:
\[
\mathbf{m}_c = C \theta \mathbf{e}_z.
\]

Finally, boundary conditions are enforced by imposing velocities at the filament endpoints (Fig. \ref{fig:parameter}). We assume the filament to be initially straight and horizontal and the proximal velocity $\dot{\mathbf{x}}_1 = (\dot{x}_1,\dot{y}_1)$ and distal velocity $\dot{\mathbf{x}}_{N+1} = (\dot{x}_{N+1},\dot{y}_{N+1})$ to follow the kinematic constraints:
\begin{equation}
\begin{array}{l l}
\dot{y}_1 (t) = 0, & \dot{y}_{N+1} (t) = 0, \\
\dot{x}_1 (t) = (1-a) V_p, & {\cmedit \dot{x}_{N+1} (t)} = -(1+a) V_p,
\end{array}
\label{eq:constraints}
\end{equation}
where $V_p$ is a given velocity and $a \in [0,1]$ is a nondimensional parameter controlling the asymmetry in the endpoint motion. The case $a = 0$ represents a fully symmetric situation: both ends move towards each other at the same speed, mimicking compression from opposing forces (e.g., cytoskeletal filaments under isotropic stress \cite{wang1998mechanical}). On the contrary, when $a=1$, the proximal end is pinned and only the distal end is moving, replicating anchored flagella or polymers in shear flow. We will see later that the velocity partition parameter $a$ influences shape transitions through elastohydrodynamic asymmetry.

Unknown contact forces $\mathbf{f}_0 = (f_{0x},f_{0y})$ and $\mathbf{f}_N = (f_{Nx},f_{Ny})$ are required at both endpoints to enforce these prescribed boundary conditions. These unknown forces can be embedded in the system of equations and solved for simultaneously with the time evolution of the $N+2$ kinematic parameters of the filament. 
Writing the balance of forces and torques over $N$ subsystems, as detailed in \cite{moreau2018asymptotic}, adding the four buckling constraints Eqs.~\eqref{eq:constraints}, and rescaling with respect to length scale $L$ and relaxation time scale $\tau = \eta_{\perp} L^4 / E_b$ -- that takes into account the flexibility of the filament against the drag of the surrounding fluid -- yields a matrix system of $N+6$ scalar ordinary differential equations:
\begin{equation}
\mathrm{Bu} \; \mathbf{A} \dot{\mathbf{X}} = \mathbf{B},
\label{eq:system}
\end{equation}
where $\mathbf{X} = \begin{pmatrix}  x_1 & y_1 & \theta &  \alpha_2 & \dots & \alpha_N & f_{0x} & f_{0y} & f_{Nx} & f_{Ny} \end{pmatrix}^T$. 
Note that the unknown $\mathbf{X}$ contains the position and shape of the filament, as well as the unknown contact forces $\mathbf{f}_0$ and $\mathbf{f}_N$, which can thus be obtained simultaneously upon resolution of System \eqref{eq:system}.
The full MATLAB code implementing the CG framework is available on GitHub \cite{hydrodynamicEulerElasticaGithub}. The dimensionless buckling number $\mathrm{Bu}$ is defined by 
$$
\mathrm{Bu} = \frac{\eta_{\perp} L^3 V_p}{E_b},
$$ 
and encapsulates the ratio between the characteristic speed at which the filament buckles ($V_p$) and the characteristic speed at which it relaxes to its equilibrium shape ($L/\tau$). Small values of $\mathrm{Bu}$ correspond to a quasistatic evolution: indeed, we then have $V_p \ll L/\tau$, giving time for the filament to approach its elastic equilibrium at all times. Its behaviour is expected to resemble the steady-state solutions of Euler-elastica. This prediction will be examined in the next section. On the other hand, a high buckling number implies $V_p \gg L/\tau$: buckling occurs much faster than relaxation, providing a snapshot of how high buckling modes emerge, propagate, and dissipate along the filament. This gives rise to complex behaviours that are considerably different from those observed in the static case.

In practice, the value of the buckling number $\mathrm{Bu}$ spans over several orders of magnitude, from floppy filaments such as flagella or long polymer chains, to rigid filaments such as short microtubules. The typical length of biological filaments lie between 1 for short microtubules \cite{kurachi1995buckling} and 250 $\mu$m for rodent sperm cells \cite{cummins1985mammalian,gaffney2011mammalian,riedel2007molecular}. The hydrodynamic drag coefficient $\eta_{\perp}$ varies between $10^{-2}$ for low viscous media and 1 Pa.s for high viscous media \cite{oriola2017nonlinear,ishimoto2018human}. While notoriously difficult to evaluate precisely at microscopic scale, bending stiffness $E_b$ may range from $1.10^{-23}$ for microtubule protein \cite{kurachi1995buckling,shin2004bending} to $2.10^{-21}$ N.m$^{2}$ for bull sperm \cite{oriola2017nonlinear}. Assuming the buckling speed $V_p$ can reasonably go from $1$ to $100$\% of the filament length per second, the dimensionless buckling number $\mathrm{Bu}$ theoretically ranges from $10^{-2}$ to $10^5$. Two representative parameter sets are summarised in Table \ref{table:1}, illustrating physical cases that belong to this numerical range.

\begin{table}
\begin{center}
\begin{tabular}{lcc}
\toprule
Parameter & Human sperm & Short microtubule \\
& high viscosity & low viscosity \\
\midrule
$L$ ($\mu$m) & 55 & 6 \\
$\eta_{\perp}$ (Pa s) & 0.14 & $5\times10^{-3}$ \\
$E_b$ (N m$^2$) & $1.2\times10^{-21}$ & $1.0\times10^{-23}$ \\
$V_p$ ($\mu$m s$^{-1}$) & 5 & 2 \\
\midrule
$\mathrm{Bu}$ & 126 & 0.21 \\
\bottomrule
\end{tabular}
\end{center}
\caption{Representative physical parameters and corresponding buckling numbers.}
\label{table:1}
\end{table}

The following section is dedicated to the results obtained through the numerical simulations \cite{hydrodynamicEulerElasticaGithub}. The equations are integrated using the \textsc{Matlab} built-in ODE solver \texttt{ode15s}. Unless otherwise stated, in all the simulations, $N=30$ and the final time is chosen equal to $0.9/\mathrm{Bu}$ (which means that the cumulative displacement by both endpoints of the filament at the end of the simulation equals $1.8 \, L$). The initial condition is given by $(x_1(0),y_1(0))=(0,0)$ and a small random perturbation for each angle, chosen according to a Gaussian distribution with mean 0 and variance $10^{-14}$. This numerically models the infinitesimal material bias that triggers buckling. 

\section{Results}

\subsection{Comparison with the Euler-elastica static case} \label{subsection:clamp}

In this section, we evaluate the influence of the buckling number $\mathrm{Bu}$ by comparing our dynamic results with the classical static case, where the filament is assumed to be in equilibrium at every instant. While low $\mathrm{Bu}$ results closely align with static predictions, increasing the buckling number triggers complex, non-equilibrium behaviours that deviate significantly from classical elastica theory.
\revision{To enable a direct comparison with classical elastica theory, we focus on the horizontal component of the contact force $f_0$ at the proximal end, $f_{0x}$, which represents the effective compressive load applied to the filament and corresponds to the axial force used in standard force--displacement characterisation of elastic rods. In particular, small values of $\mathrm{Bu}$ correspond to a quasi-static regime, in which the filament has sufficient time to relax towards elastic equilibrium at each instant. In this limit, the dynamics closely follow the classical Euler--elastica solutions. In contrast, for large $\mathrm{Bu}$, the filament is driven faster than it can relax, leading to force--displacement responses that have no direct static analogue.}

\begin{figure}
\begin{center}
\includegraphics[width=0.45\textwidth]{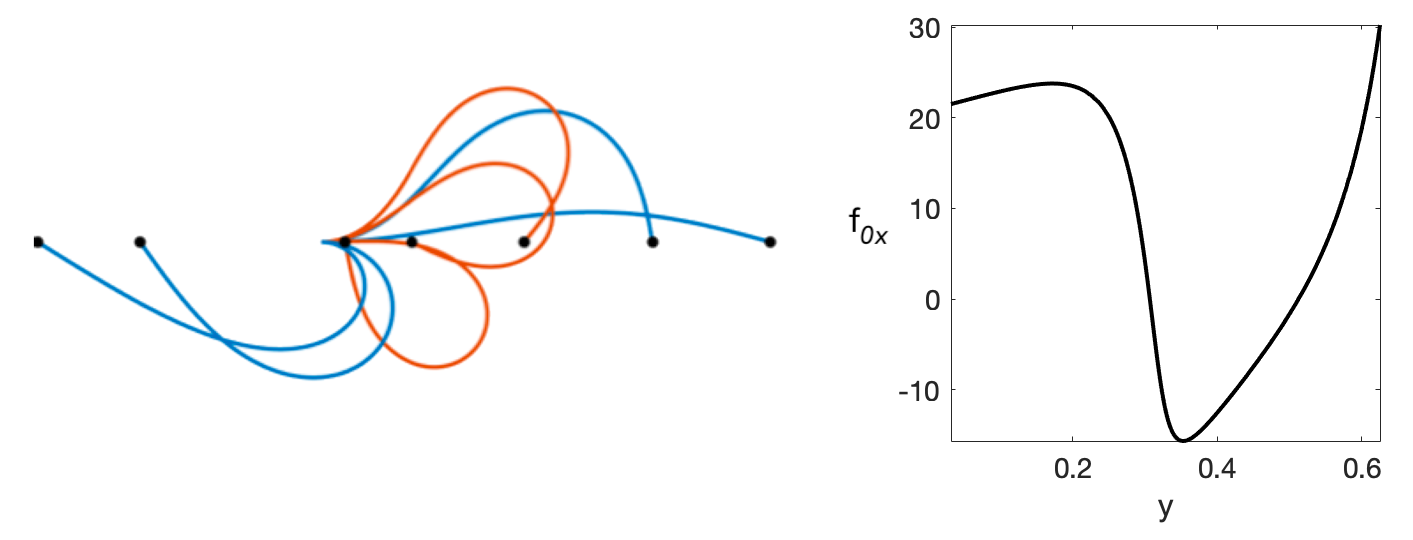}
\caption{Solution of static Euler-elastica in the clamped case, as represented in \cite{gadelha2018filament}, with the evolution of the horizontal force $f_{0x}$ with respect to displacement $y$ plotted on the right. The filament shape is displayed at selected times, in blue (resp. red) when $f_{0x}$ is positive (resp. negative), denoted as the ``flipping'' phase here.}
\label{fig:elastica}
\end{center}
\end{figure}

In this static case, the shape of the filament, continuously parameterised by its arclength $s \in [0,L]$, angle $\theta$, bending stiffness $E_b$ and force magnitude $F$ at the endpoints is well-known to satisfy the following equation:
\[ E_b \theta_{ss} + F \sin \theta = 0, \]
with given positional conditions at the endpoints. The solution of this problem, called the Euler-elastica, can be analytically derived and has been widely studied \cite{fung2017classical,landau1986course,timoshenko2009theory,book:antman,gadelha2018filament}. In particular, a study of the shapes adopted by the filament for clamped and pinned endpoints has been conducted in \cite{gadelha2018filament} (which mainly considers a filament-bundle model). As an example of a static buckling solution, Fig.~\ref{fig:elastica} shows the aspect of the filament with a clamped extremity experiencing buckling, as computed in \cite{gadelha2018filament}, as well as the evolution of the horizontal force at the clamped end. 

\begin{figure*}
\begin{center}
\includegraphics[width=\textwidth]{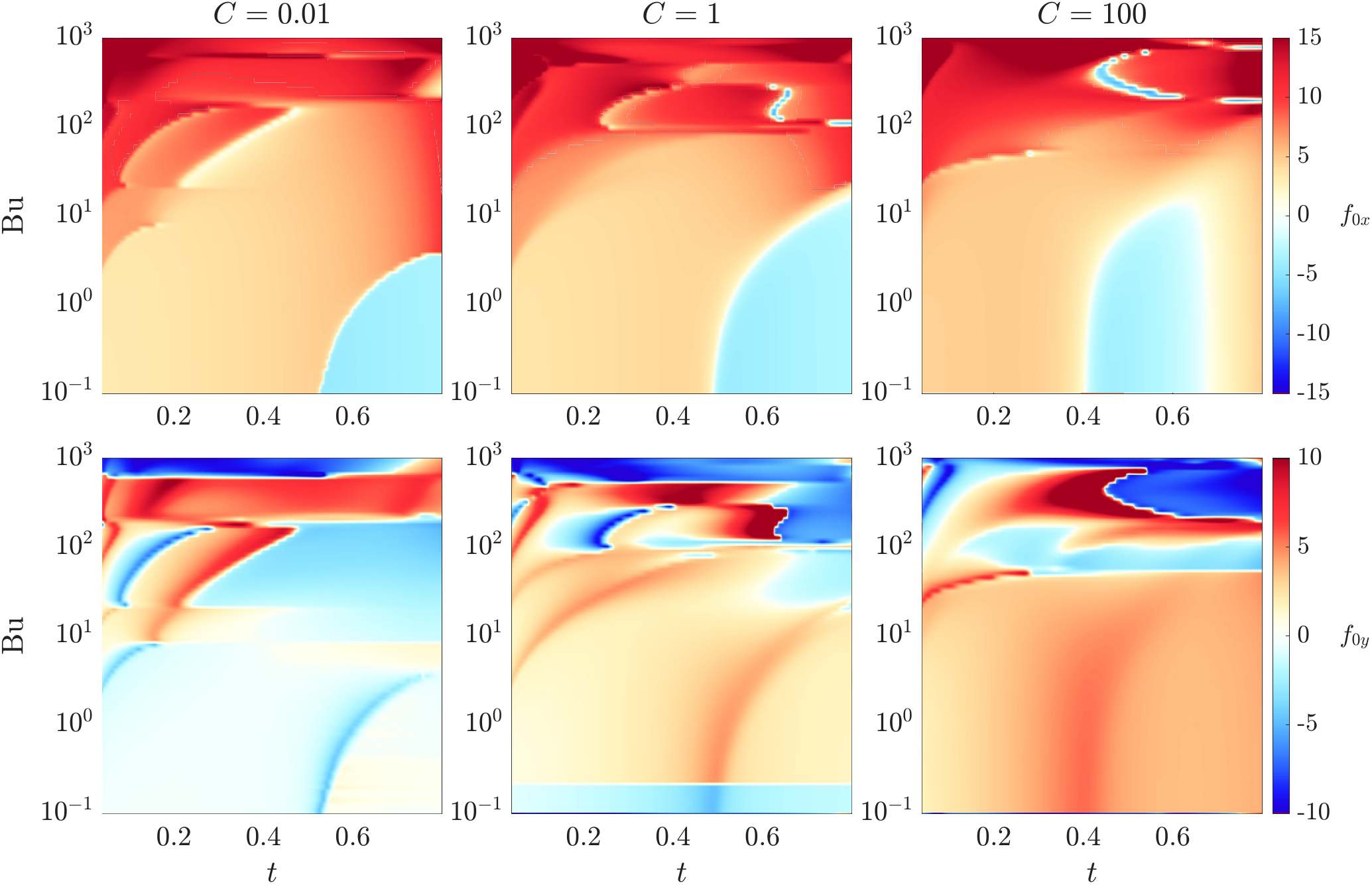}
\caption{{\cmedit Force-response maps for three clamping strengths, $C=10^{-2}$, $1$ and $100$. The top row shows the function of the horizontal force $f_{0x}$ given by $\operatorname{sgn}(f_{0x})|f_{0x}|^{1/2}$ for better visualisation, while the bottom row shows the corresponding quantity for $f_{0y}$. Both are plotted as functions of nondimensional time $t$ and buckling number $\mathrm{Bu}$. Low $\mathrm{Bu}$ responses are smooth and consistent with quasi-static force evolution, whereas high $\mathrm{Bu}$ responses develop strongly structured force bands associated with transient multimode excitation.}}
\label{fig:clamp}
\end{center}
\end{figure*}

\revision{
For the strongly clamped case, the low-$\mathrm{Bu}$ horizontal force response closely follows the static Euler--elastica curve shown in Fig.~\ref{fig:elastica}. This agreement confirms that, when endpoint motion is slow compared with elastohydrodynamic relaxation, the dynamic model recovers the quasi-static elastica limit. Figure~\ref{fig:clamp} then extends the comparison across buckling number and clamping strength, showing how this quasi-static correspondence breaks down as $\mathrm{Bu}$ increases.
}
{\cmedit
Figure~\ref{fig:clamp} extends this comparison across buckling number and clamping strength. Rather than plotting a single force--displacement curve, it shows force-response maps over the full simulated range of $\mathrm{Bu}$ for three representative clamping strengths, from nearly pinned ($C=10^{-2}$) to nearly clamped ($C=100$). The low-$\mathrm{Bu}$ region remains smooth and consistent with the quasi-static force evolution associated with flipping, especially under strong clamping, whereas increasing $\mathrm{Bu}$ produces broad regions of rapidly varying axial force response that have no direct Euler-elastica counterpart. The transverse component $f_{0y}$, shown in the bottom row, is small and smooth at low $\mathrm{Bu}$ but becomes strongly structured at high $\mathrm{Bu}$, reflecting the growth of asymmetric transient deformations.

The maps also show that the high-$\mathrm{Bu}$ response is only partially controlled by the clamping condition. Changing $C$ shifts and reshapes the force-response regions, but the onset of rapidly varying bands at large $\mathrm{Bu}$ persists across all three columns. This suggests that the early excitation and subsequent decay of higher curvature modes is not merely a boundary-torque effect, motivating the multimode analysis developed in Section~\ref{subsection:stability}.
}

Having established that dynamic buckling strongly influences the filament force response and its departure from the static Euler-elastica limit (compare Fig.~\ref{fig:elastica} with the low- and high-$\mathrm{Bu}$ regions of Fig.~\ref{fig:clamp}), we now seek to characterise the distinct shape regimes that emerge as the buckling number $\mathrm{Bu}$ is varied.

\subsection{Buckling-number-dependent shape selection: flip, loop and knot, with coexistence \label{section:buckling}}

\begin{figure*}
\begin{center}
\includegraphics[width=0.95\textwidth]{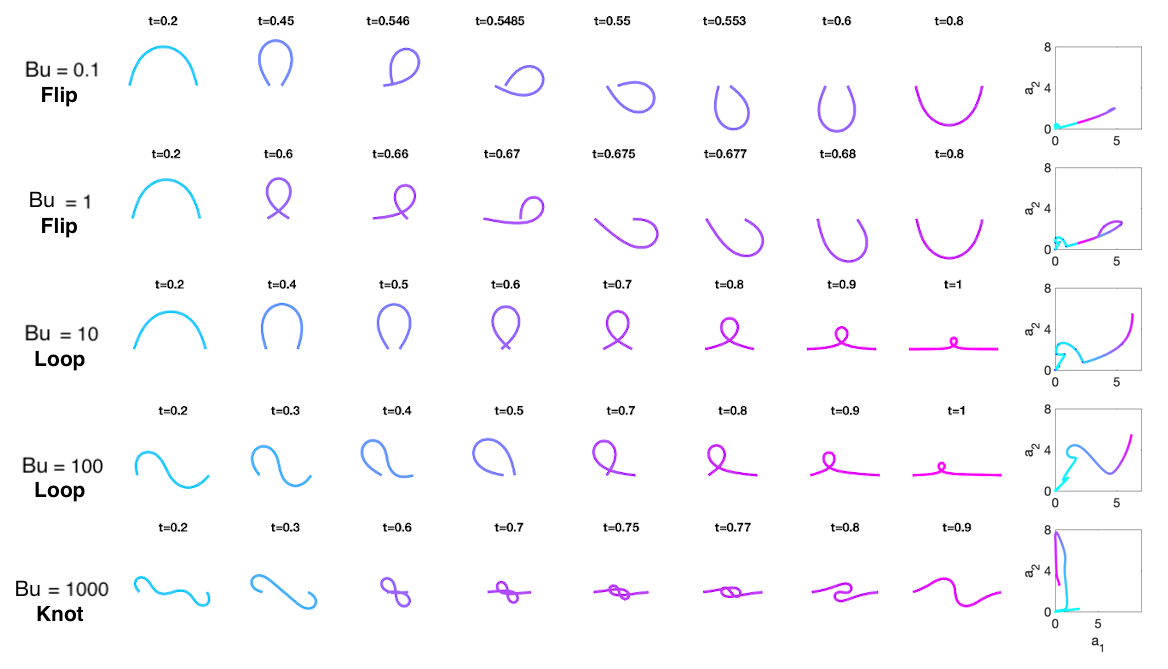}
\caption{Filament shapes at different times and buckling numbers $\mathrm{Bu}$. The first two rows show flipping at low speed, the middle rows show loop formation at medium-range $\mathrm{Bu}$, and the last row shows a knot-like self-intersecting configuration at high $\mathrm{Bu}$. 
The column of panels on the right displays each of the five trajectories in the ``Fourier plane'' $(a_1,a_2)$: the amplitudes of the first and second spatial Fourier modes of the filament curvature are plotted on the $x$ and $y$ axes, respectively. The coloured line indicates the time evolution, from light blue at initial time to pink at the end of the simulation.}
\label{fig:snaps}
\end{center}
\end{figure*}

\begin{figure*}
\begin{center}
\includegraphics[width=0.58\textwidth]{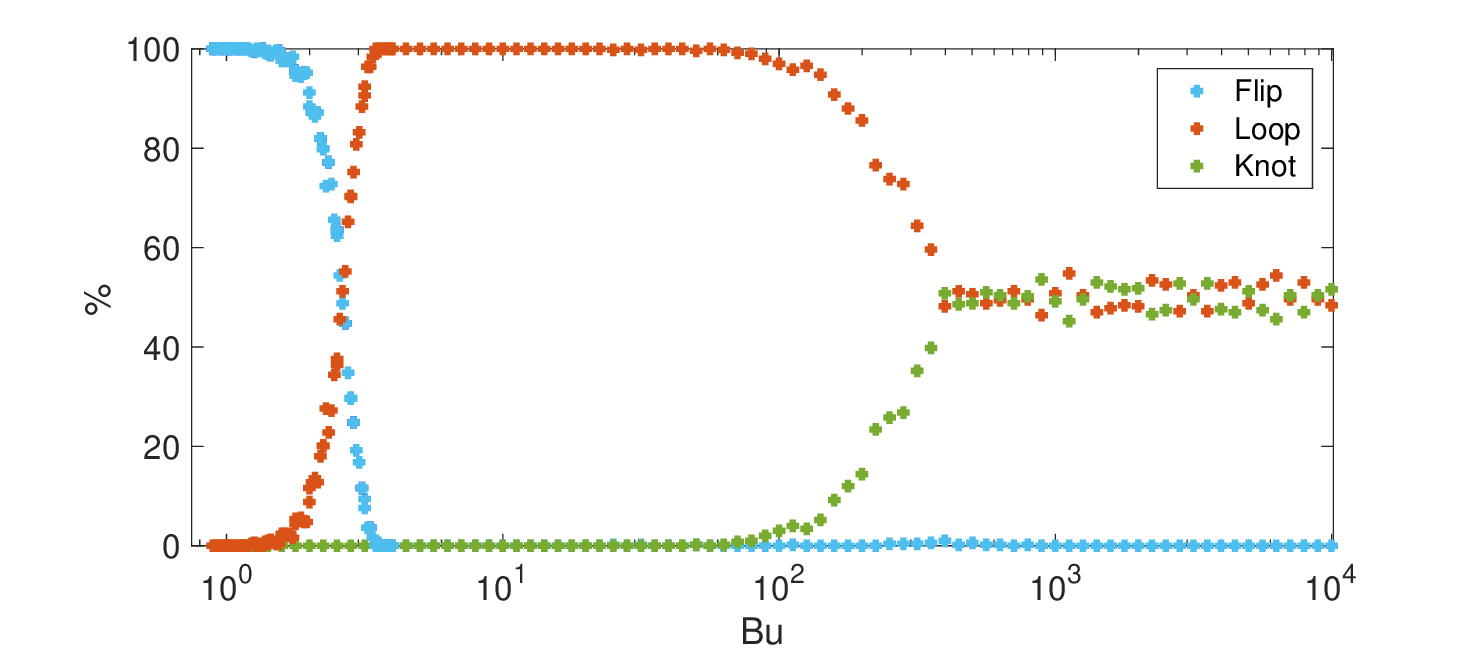}
\includegraphics[width=0.31\textwidth]{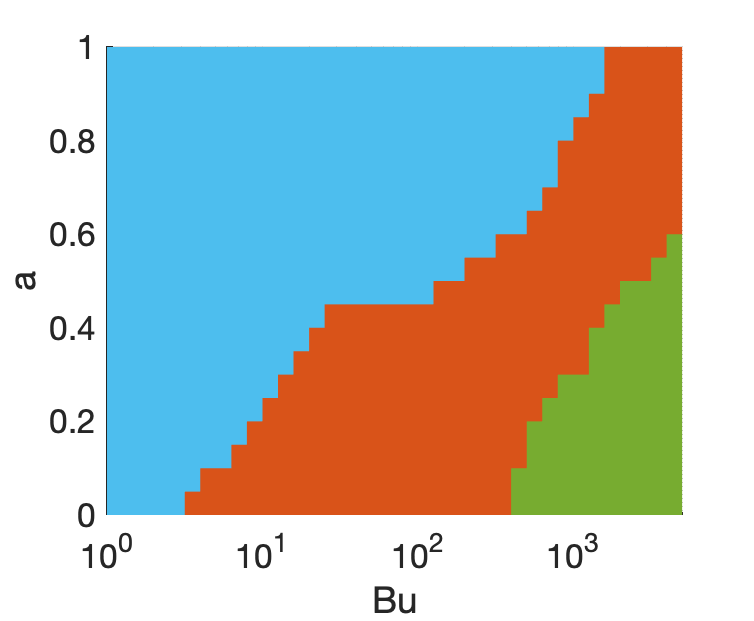}
\caption{Predominance of each shape depending on nondimensional speed $\mathrm{Bu}$ and asymmetry parameter $a$. For every value of $\mathrm{Bu}$ and $a$, 500 simulations were run and the final-time shape was classified through Fourier analysis as a flip, loop or knot. The left panel shows the frequency of appearance of these shapes with respect to $\mathrm{Bu}$, for $a=0$. The right panel shows the regions where the flip, loop and loop/knot coexistence appear with respect to $\mathrm{Bu}$ and $a$, respectively in blue, red and green.}
\label{fig:shapes}
\end{center}
\end{figure*}

In this section, we focus on the pinned case ($C=0$), where the departure from the static response is particularly pronounced (see the low-clamping force maps in Fig.~\ref{fig:clamp}), and analyse the behaviours that emerge as the buckling number $\mathrm{Bu}$ is varied. By computing the shape evolution across a wide range of $\mathrm{Bu}$, we find that the dynamics fall into three main categories: the `flip', the `loop' and the `knot' (Fig.~\ref{fig:snaps}). For relatively low $\mathrm{Bu}$ ($10^{-1}$ to $1$), we observe the flip behaviour: a loop forms as the endpoints cross, but it is unstable and the filament quickly flips, recovering a U-shape at long times. As $\mathrm{Bu}$ increases ($10^{1}$ to $10^{2}$), the loop stabilises and persists as the final configuration.

At high $\mathrm{Bu}$ (approximately $\geqslant 10^3$), the filament develops self-intersecting configurations, with multiple crossings occurring before partial untangling. As a slight abuse of language, we refer to this configuration as a `knot'. This regime reflects rapid buckling that generates high-curvature intersecting regions, akin to the `writhing' observed in 3D flagella \cite{lough2023self}. We emphasise that these are not true knots: they arise here because the model is planar and does not include steric interactions to prevent self-contact (in the spirit of Euler--elastica solutions). The same caveat applies to the looped configurations. Nevertheless, their emergence signals a regime in which elastic and hydrodynamic forces compete strongly, as we discuss in the following section.

The spatial discrete Fourier transform $(A_1, \dots, A_{N-1})$ of the filament curvature vector, described by the angles $(\alpha_2, \dots , \alpha_N)$, constitutes a convenient tool to identify and classify these shapes. Indeed, we observed that the evolution of the amplitudes $a_1 = |A_1|$ and $a_2 = |A_2|$ of the first two modes in the plane $(a_1,a_2)$ clearly separates the three different shapes; see the right column of panels in Fig.~\ref{fig:snaps}. The first two panels (low $\mathrm{Bu}$) display a returning branch and overlapping that characterise the flip case. The profiles of the third and fourth panels indicate a final loop shape. The last panel shows a marked change of behaviour at high $\mathrm{Bu}$ for the knot shape, including the near-vanishing of the first mode; note the change of scale on the x-axis. In particular, the amplitude $a_1$ of the first mode is significantly different at the end of the simulation for the three shapes. 

To highlight the importance of $\mathrm{Bu}$ in filament shape selection, we measured the frequency of appearance of each shape over $500$ simulations for $\mathrm{Bu}$ ranging from $0.5$ to $10^4$.
The left panel of Fig.~\ref{fig:shapes} displays the results, showing the shape predominance with respect to $\mathrm{Bu}$. A rather surprising observation is that, for a fixed value of $\mathrm{Bu}$, the filament may adopt one or another of the three behaviours, depending on the initial condition -- no matter how small the random bias in the initial shape is set. A coexistence between the flip and loop shapes occurs for a narrow range of buckling numbers, roughly between $\mathrm{Bu}=1.5$ and $\mathrm{Bu}=2.5$. For $\mathrm{Bu} < 1.5$, the filament always flips; for $\mathrm{Bu}>2.5$, it always loops, until $\mathrm{Bu}$ reaches approximately $3\times 10^3$. {\cmedit The representative snapshots in Fig.~\ref{fig:snaps} show that self-intersecting knot-like trajectories can already be produced around $\mathrm{Bu}=10^3$, whereas the statistical phase diagram indicates that they become frequent only at larger $\mathrm{Bu}$.} Above this range, the filament sometimes adopts a knot shape; and the loop and knot shapes become seemingly equiprobable for very high values of $\mathrm{Bu}$. 

We also investigated the role of the symmetry parameter $a$ (see Eq.~\eqref{eq:constraints}) in the filament buckling behaviour. Along with the buckling number $\mathrm{Bu}$, we observed that $a$ plays an important role in the appearance of the three outcomes. The regions of predominance of the three shapes with respect to $\mathrm{Bu}$ and $a$ are displayed on the right panel of Fig.~\ref{fig:shapes}: blue for the flip, red for the loop and green for coexistence between loop and knot. The buckling number range for which the loop shape predominates notably shifts towards high values of $\mathrm{Bu}$ as $a$ increases; similarly, the unstable knot shape only starts appearing for higher buckling numbers, and even completely disappears for high values of $a$. This can be interpreted in the following way: when $a>0$, the induced asymmetry in the endpoint motion creates a hydrodynamic effect that drags the filament to the left or right, making it easier for the loop to untie and flip, like on the second row of panel (a).

Further investigation in Section \ref{subsection:stability} shows that the final shape for the ranges of buckling number where two shapes coexist is highly unpredictable, despite the model being deterministic and the initial conditions being extremely close to each other.

\subsection{Fourier-mode dynamics of filament curvature}
\label{subsection:stability}
\revision{
To characterise the deformation of the filament during compression, we analyse the curvature field $\kappa(s,t)$ along the arc length using a Fourier decomposition,
\begin{equation}
\kappa(s,t)=\sum_{k\ge1} a_k(t)\sin\!\left(\frac{k\pi s}{L}\right),
\end{equation}
where $a_k(t)$ denotes the amplitude of the $k$-th curvature mode.
In the simulations, this modal decomposition is evaluated on the discrete curvature vector $(\alpha_2,\dots,\alpha_N)$ using its discrete Fourier transform. We therefore use $A_k$ for the discrete complex Fourier coefficient and $a_k=|A_k|$ for its amplitude, which is the quantity plotted throughout.
This representation provides a convenient way to quantify the contribution of different spatial scales to the filament shape. 
The temporal evolution of the modal amplitudes $a_k$ is shown in Fig.~\ref{fig:decay}(a--c). 
{\cmedit The time variable in Fig.~\ref{fig:decay} is the same nondimensional simulation time used to follow the early compression dynamics; it should therefore be read as the early-time modal counterpart of the later shape snapshots and Fourier trajectories shown in Figs.~\ref{fig:snaps} and~\ref{fig:3d}.}
The simulations reveal a clear ordering of the modal dynamics: several modes are excited during the initial stages of buckling, but higher-order modes decay rapidly, while the long-time behaviour is dominated by the first few modes. 
In particular, each mode exhibits a well-defined transient maximum before relaxing, and the time at which this maximum occurs decreases markedly with increasing mode number.

To rationalise these observations, we consider a simplified linear elastohydrodynamic balance capturing the competition between compression and bending relaxation. 
Under the small-slope approximation, where the curvature may be approximated as $\kappa \approx \partial_s^2 y$, the overdamped dynamics leads to an evolution equation for the curvature field,
\begin{equation}
\eta_\perp \partial_t \kappa
=
- E_b \partial_s^4 \kappa - F \partial_s^2 \kappa ,
\end{equation}
where $E_b$ is the bending rigidity, $\eta_\perp$ the transverse viscous drag coefficient, and $F$ denotes an effective compressive load associated with the imposed end displacement. 
Projecting this equation onto the sinusoidal modes defined above yields the modal evolution equation
\begin{equation}
\dot a_k=\lambda_k a_k,
\qquad
\lambda_k=\frac{1}{\eta_\perp}
\left(
F\frac{k^2\pi^2}{L^2}
-
E_b\frac{k^4\pi^4}{L^4}
\right).
\end{equation}
This expression highlights the competition between compression-driven amplification, which scales as $k^2$, and bending-induced damping, which increases as $k^4$. 
As a consequence, short-wavelength curvature modes experience much stronger elastic regularisation and therefore relax much more rapidly than lower-order modes. 
This strong $k^4$ dependence explains the rapid disappearance of high-$k$ contributions observed in Fig.~\ref{fig:decay}(a--c).

While this linear relaxation mechanism accounts for the rapid damping of higher modes, the simulations show that each mode first grows before reaching a transient maximum. 
To capture this behaviour, we introduce a minimal excitation--relaxation model for the modal amplitudes,
\begin{equation}
\dot a_k = \lambda_k a_k + f_k e^{-t/\tau_f},
\end{equation}
where the second term represents a short-lived forcing associated with the initial compression transient generated by the imposed boundary motion. 
Here $f_k$ denotes the amplitude of the transient excitation acting on mode $k$, and $\tau_f$ is the characteristic duration of the forcing phase. 
In this framework, the transient peak observed in Fig.~\ref{fig:decay}(a--c) arises from the competition between early-time excitation and elastohydrodynamic relaxation. 
Because the relaxation rate increases rapidly with the mode number through the bending term, higher-order modes reach their maximum amplitude earlier than lower-order modes.

This ordering of peak times is quantified in Fig.~\ref{fig:decay}(d), which shows the dependence of the peak time $T_k^{\max}$ on the mode number. 
Over the range of modes resolved in our simulations, the peak times are well described by the empirical relation
\begin{equation}
T_k^{\max}=T_0 e^{-\beta k^{1/4}},
\end{equation}
where $T_0$ and $\beta$ are fitting parameters that depend on the buckling number. 
{\cmedit Direct fits to the high-mode range $k=4,\dots,12$ give $\beta \approx 10$, remaining of the same order across the simulated buckling numbers.}
We emphasise that this relation should be regarded as a phenomenological description of the observed hierarchy of modal peak times rather than as a universal asymptotic scaling. 
Its rapid decrease with $k$ is nevertheless consistent with the strong elastohydrodynamic damping of short-wavelength bending modes predicted by the linear argument above.

The rapid decay of higher-order modes implies that the long-time dynamics of the filament can be effectively described using only a few low-order Fourier amplitudes. In particular, the predominant role of the first three modes $a_1$, $a_2$ and $a_3$ confirms that representing the trajectories in a low-dimensional ``Fourier space'' $(a_1,a_2)$ or $(a_1,a_2,a_3)$ is relevant for visualising and analysing their behaviour.

Figure \ref{fig:3d} displays the evolution of these Fourier mode amplitudes for different values of $\mathrm{Bu}$ and twenty trajectory realisations. Panels (a) and (b) illustrate the dynamics for values of $\mathrm{Bu}$ for which the flip and loop shapes coexist (roughly $1.5 < \mathrm{Bu} < 2.5$). All the trajectories initially follow a common path before bifurcating either towards the flip or towards the loop shape state. Note that the bifurcation towards the flip occurs later and more rarely for $\mathrm{Bu}=2.1$ than for $\mathrm{Bu}=1.6$, indicating that the loop shape becomes more predominant. The fact that these trajectories diverge from a single path emphasises the sensitivity of the system to small perturbations: infinitesimal differences in the early dynamics are amplified as the higher modes decay, eventually pushing the filament into different shape states.

Panel (c) shows the evolution of the first and second Fourier mode amplitudes when $\mathrm{Bu}=200$. For this value, all trajectories eventually end up as loop shapes. Nonetheless, they follow a wide range of paths in the Fourier amplitude phase-space before reaching this final state, illustrating the complex transient dynamics the filament may undergo. Panel (e) additionally displays the third mode $a_3$, highlighting how the first three Fourier amplitudes evolve jointly in time.

Finally, panels (d)--(f) show the emergence of knots at high buckling numbers ($\mathrm{Bu}=1000$). As shown in Fig.~\ref{fig:shapes} and discussed above, in this regime the filament ends up equiprobably as a loop or as a knot. The knot trajectories are visible on the left-hand side of graph (d) and are characterised by a strongly attenuated first mode. Panel (f) shows how these knot trajectories evolve according to a complex pattern in the $(a_2,a_3)$ plane. Of particular interest is that the two families of trajectories (loops and knots) can already be distinguished at very early times. This suggests that higher-order modes, despite their rapid decay, may still influence the early dynamics and contribute to determining the final outcome.
}

\bd{
\begin{figure*}
\begin{center}
\includegraphics[width=\textwidth]{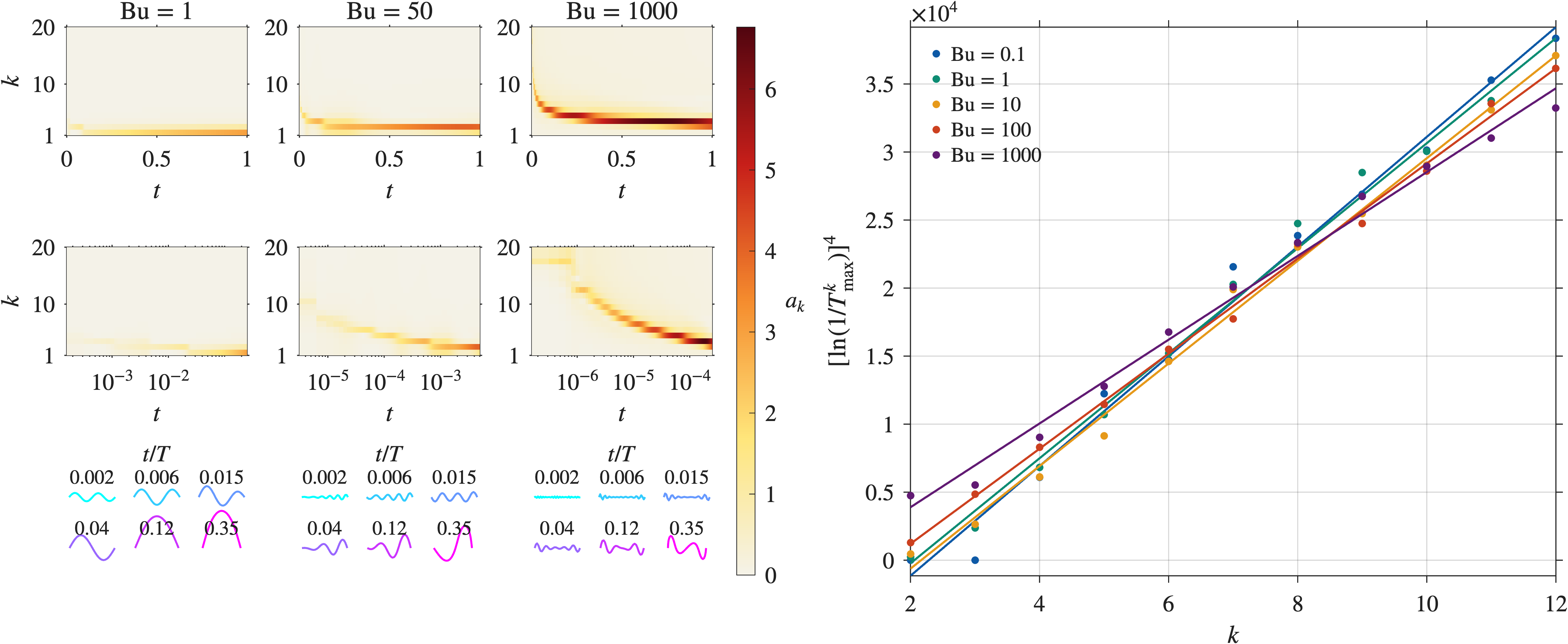}
\caption{Early-time evolution and peak-time scaling of curvature Fourier modes. Heatmaps show the amplitudes $a_k=|A_k|$ of the first $20$ curvature Fourier modes ($k=1,\dots,20$; vertical axis) as a function of time (horizontal axis) for three representative buckling numbers, $\mathrm{Bu}=1,50,1000$ (white: low amplitude; dark red: high amplitude). The top row uses a linear time axis $t$, while the middle row displays the same data against $\log(t)$, highlighting an early-time cascade in which higher-$k$ modes peak earlier and decay rapidly. The bottom row shows early-time snapshots with exaggerated $y$-axis, for visual comparison. {\cmedit Right: collapse of the peak times $T_{\max}^k$ across $\mathrm{Bu}$, plotted as $[\ln(1/T_{\max}^k)]^4$ versus $k$ as in the figure. Direct fits of $T_k^{\max}=T_0\exp(-\beta k^{1/4})$ over $k=4,\dots,12$ give $(\beta,T_0)=(10.8,3.9\times10^2),(10.6,2.8\times10^2),(11.1,7.4\times10^2),(9.5,4.6\times10^1),(8.4,6.2)$ for $\mathrm{Bu}=0.1,1,10,100,1000$, respectively.}}
\label{fig:decay}
\end{center}
\end{figure*}

\begin{figure*}[t]
\begin{center}
\includegraphics[width=\textwidth]{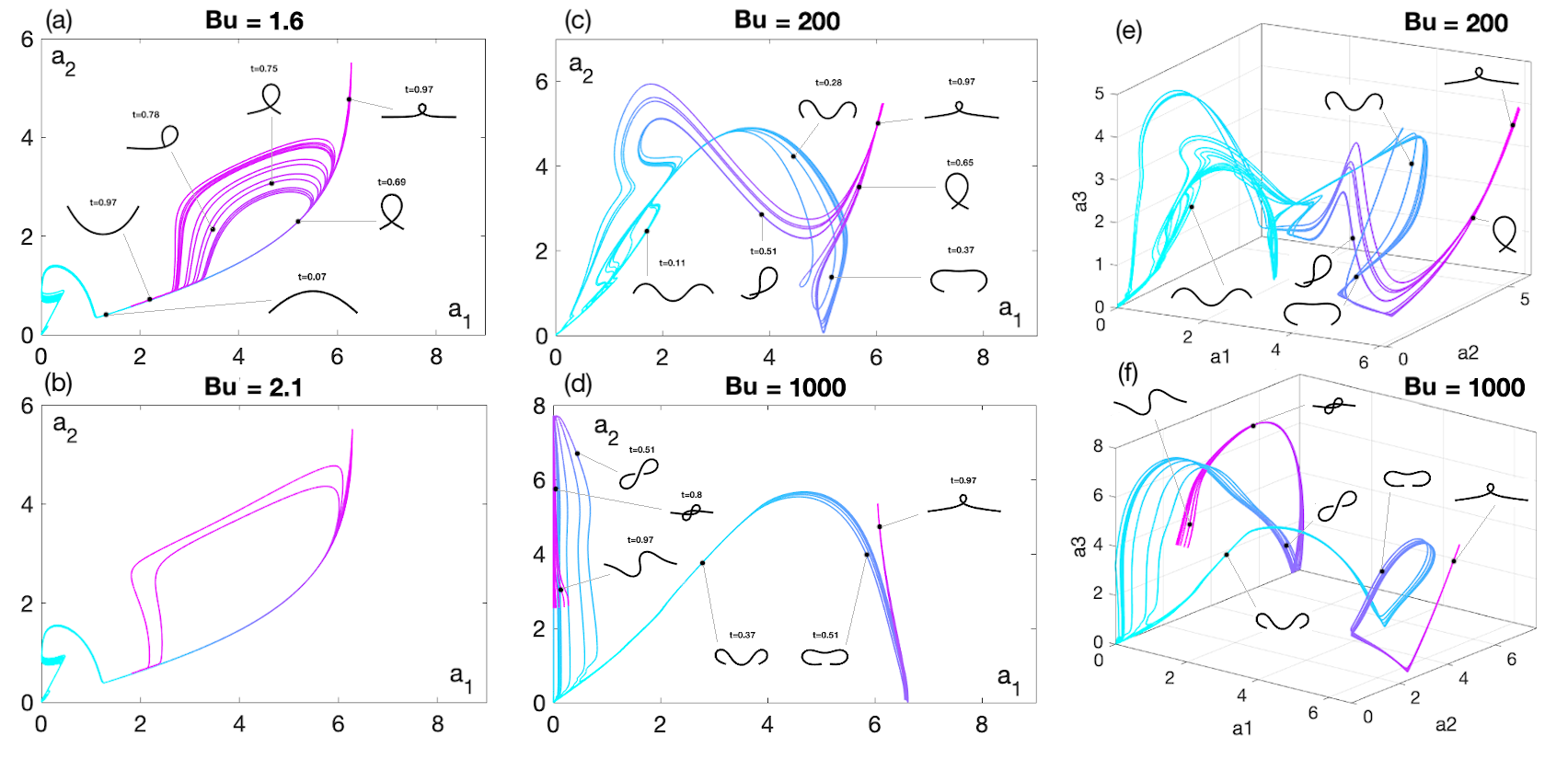}
\caption{Bifurcation and instability phenomena across buckling regimes. For each panel (a)--(f), $20$ trajectories are shown. Panels (a)--(d) plot $(a_1,a_2)$, the amplitudes of the first two curvature Fourier modes, on the $x$- and $y$-axes, respectively; panels (e) and (f) additionally include $a_3$ on the $z$-axis. Trajectories are coloured from light blue (initial time) to magenta (final time); all start at the origin since the initial configuration is straight and all Fourier modes vanish. Representative filament shapes at selected instants are overlaid in (a), (c) and (d). Time is normalised so that the final time equals $1$ in every panel. Panels (a) and (b) lie in the range of $\mathrm{Bu}$ where the flip and loop outcomes coexist, while panels (c)--(f) illustrate the modal dynamics at larger $\mathrm{Bu}$.}

\label{fig:3d}
\end{center}
\end{figure*}
}

\section{Conclusion}

In this paper, we presented an exploratory numerical investigation into the nonlinear buckling dynamics of elastic microfilaments undergoing compression. By leveraging the asymptotic framework developed in \cite{moreau2018asymptotic}, we extended the classical study of buckling beyond the quasi-static equilibrium, revealing a rich landscape of time-dependent morphologies that emerge when hydrodynamic drag competes with elastic relaxation.

Our findings demonstrate that the long-time evolution of a buckled filament is primarily governed by the dimensionless buckling number $\mathrm{Bu}$, which measures the ratio between the characteristic timescale of elastic relaxation and the timescale of imposed deformation. We have shown that $\mathrm{Bu}$ organises three distinct morphological regimes: an elasticity-dominated regime in which the filament flips back to a simple U-shaped configuration, a viscous-stabilised regime in which it remains in a looped configuration, and a high-$\mathrm{Bu}$ regime characterised by multimode transients and knot-like self-intersections.

Crucially, we identified a ``coexistence zone'' where the system exhibits extreme sensitivity to initial conditions. In this regime, nearly identical starting configurations bifurcate into fundamentally different topological outcomes, suggesting that deterministic microfilament buckling can lead to stochastic-like morphological diversity.

The application of spatial Fourier transforms to the filament curvature provided a rigorous metric for these transitions. Our analysis revealed a ``spectral filtering'' process: while high-order modes are excited during the initial transient phase of buckling, the long-term behaviour is dictated by the interaction of only the first two or three modes. This reduction in effective degrees of freedom suggests that the complex dynamics of biological filaments might be captured by low-dimensional reduced-order models.

This study opens several avenues for further research. First, a formal theoretical stability analysis of the governing equations is required to analytically bound the transition regions between the flip, loop, and knot regimes. Such an analysis would provide a more rigorous basis for the numerical boundaries observed in our phase diagrams. Furthermore, while our 2D model highlights the emergence of ``pseudo-knots'', many biological filaments operate in a three-dimensional environment. Extending this framework to 3D would allow for the exploration of writhing, twisting, and true topological knotting, which are critical to understanding the propulsion of flagellated microorganisms and the packing of DNA \cite{cammann2025form,du2019dynamics}. Ultimately, the sensitivity of the buckling outcome to the parameter $a$ suggests that biological systems and artificial robotic platforms may actively tune boundary asymmetries to control filament reconfiguration.

\section*{Acknowledgments}
This work was supported by the Agence Nationale
de la Recherche (Grants No. ANR-21-CE45-0013).

\bibliographystyle{apsrev4-2}
\bibliography{biblio_elasto}

\end{document}